\documentclass{PoS}

\usepackage{microtype}

\newcommand{\tb}{\bar t}

\newcommand{\ttbh}{ t \tb H}
\newcommand{\al}{\alpha}
\newcommand{\als}{\alpha_{\rm s}}
\newcommand{\shat}{\hat s}
\newcommand{\sigh}{\hat \sigma}
\newcommand{\si}{\sigma}
\newcommand{\nn}{\nonumber}
\newcommand{\tosv}{{\scriptscriptstyle \to}}
\newcommand{\CF}{C_{\mathrm{F}}}
\newcommand{\CA}{C_{\mathrm{A}}}

\title{
	\vspace{-3em}\begin{flushright}\normalsize \sf MS 16-24\\[3em]\end{flushright}
	Soft gluon resummation at fixed invariant mass for associated $t\bar{t}H$ production at the LHC}

\ShortTitle{Soft gluon resummation at fixed invariant mass for associated $t\bar{t}H$ production at the LHC} 

\author{Anna Kulesza\\
        Institute for Theoretical Physics, WWU M\"unster, D-48149 M\"unster, Germany\\
        E-mail: \email{anna.kulesza@uni-muenster.de}}

\author{Leszek Motyka, Tomasz Stebel\\
          Institute of Physics, Jagellonian University, S. \L{}ojasiewicza 11, 30-348 Krak\'ow, Poland\\
        E-mail: \email{leszek.motyka@uj.edu.pl}, \email{tomasz.stebel@uj.edu.pl}}

\author{\speaker{Vincent Theeuwes}\\
          Department of Physics, SUNY Buffalo, 261 Fronczak Hall, Buffalo, NY 14260-1500, USA\\
        E-mail: \email{vtheeuwe@buffalo.edu}}

\abstract{In the following we present our results on resummation of invariant mass threshold corrections for the $2 \to 3$ type hadronic production processes in the Mellin moment space formalism. This method is applied to the associated Higgs boson production process $pp \to t \bar{t} H$ at the LHC. The results for the total cross section, the differential distribution with respect to the invariant mass and their uncertainties are presented.}

\FullConference{Fourth Annual Large Hadron Collider Physics\\
	13-18 June 2016\\
	Lund, Sweden}

\begin{document}

\section{Introduction}
 One of the main tasks of the current LHC run is to establish the properties of the Higgs boson discovered at the LHC in 2012~\cite{Aad:2012tfa, Chatrchyan:2012ufa}. The production process in association with top quarks, $pp \to \ttbh$, provides  a direct way to probe the strength of the top Yukawa coupling without making any assumptions regarding its nature. This necessitates an improvement of the theoretical accuracy with which theoretical predictions for $pp \to \ttbh$ are known. A great amount of progress has been achieved in the recent years in this field. Although the next-to-leading-order (NLO) QCD, i.e. ${\cal O}(\als^3\al)$ predictions are already known for some time ~\cite{Beenakker:2001rj, Reina:2001sf}, they have been newly recalculated and matched to parton showers in~\cite{Hirschi:2011pa, Frederix:2011zi, Garzelli:2011vp, Hartanto:2015uka}. As of late, the mixed QCD-weak corrections~\cite{Frixione:2014qaa} and QCD-EW corrections~\cite{Yu:2014cka, Frixione:2015zaa} of ${\cal O}(\als^2\al^2)$, as well as the NLO QCD corrections to the hadronic $\ttbh$ production with top and antitop quarks decaying into bottom quarks and leptons~\cite{Denner:2015yca} are also available. However, calculations of the next-to-next-to-leading-order (NNLO) QCD corrections are currently technically out of reach. It is nevertheless interesting to ask the question what is the size and the effect of certain classes of QCD corrections of higher than NLO accuracy. One type of contribution is from soft gluon emission in the threshold limit. In this context the absolute threshold corrections have been included to all orders in perturbation theory~\cite{Kulesza:2015vda}. Also in the invariant mass threshold limit the approximation of the NNLO QCD corrections has been obtained~\cite{Broggio:2015lya}. In this contribution we report on the inclusion of the contributions from soft gluon emission in the invariant mass threshold limit to all orders in perturbation theory.

The traditional (Mellin-space) resummation formalism which is often applied in this type of calculations has been very well developed and copiously employed for description of the $2 \to 2$ type processes at the Born level. The universality of resummation concepts warrants their applications to scattering processes with many partons in the final state, as shown in a general analytical treatment developed for arbitrary number of partons~\cite{Bonciani:2003nt, Aybat:2006wq}. In particular, using a concept of individual weights for each of the functions describing different type of dynamics, be it hard, soft/collinear or soft, the factorization of the cross sections into these functions can be shown~\cite{Contopanagos:1996nh}. At the level of a specific process, adding one more particle or a jet in the final state requires accounting for  more complicated kinematics and a possible change in the colour structure of the underlying hard scattering. In the general framework the former will manifest itself in the appearance of new type of weights, strictly related to the definition of a considered observable, while the latter influences the soft and hard functions. More specifically, for processes with more than three partons involved at the Born level, the non-trivial colour flow influences the contributions from wide-angle soft gluon emissions which have to be included at the next-to-leading-logarithmic (NLL) accuracy. The evolution of the colour exchange at NLL is governed by the one-loop soft anomalous dimension which then needs to be calculated.

In the following we discuss these modifications for a generic $ ij \to kl B$ process, where $i,j$ denote massless coloured partons, $k, l$ are two massive coloured particles and $B$ is a massive colour-singlet particle. The corrections are considered in the limit of invariant mass threshold with the corresponding weight given by $z_5 = 1-(p_k+p_l+p_B)^2/\hat{s}$. Subsequently we apply the results to the case of the associated Higgs boson production with top quarks, where in the threshold limit the cross section receives enhancements in the form of logarithmic corrections in $z_5$, i.e. $(\log^i z_5/z_5)_+,\;i=0,1,..$. The quantity $z_5$ measures the fraction of the total initial state energy that goes into the gluon emission.  An additional improvement of the calculation at the NLL accuracy is achieved by including the ${\cal O}(\als)$ non-logarithmic threshold corrections originating from hard dynamics.

\section{Resummation at production threshold}
At the partonic level, the Mellin moments for the process $ ij \to kl B$ are given by
\begin{equation}
\frac{d\sigh_{ij \to kl B, N}}{dQ^2} (m_k, m_l, m_B, \mu_F^2, \mu_R^2) = \int_0^1 d \hat\rho \, \hat\rho^{N-1} \frac{d\sigh_{ij \to kl B}}{dQ^2} (\hat \rho, m_k, m_l, m_B, \mu_F^2, \mu_R^2) 
\end{equation}
with $\hat \rho = 1- z_5=Q^2/\shat$, $Q^2=(p_l+p_k+p_B)^2$.

At LO, the $\ttbh$ production receives contributions from the $q \bar q$ and $gg$ channels. We analyze the colour structure of the underlying processes in the $s$-channel colour bases, $\{ c_I^q\}$ and $\{c_I^g\}$, with \mbox{$c_{\bf 1}^q =  \delta^{\al_{i}\al_{j}} \delta^{\al_{k}\al_{l}},$} 
\mbox{$c_{\bf 8}^{q} = T^a_{\al_{i}\al_{j}} T^a_{\al_{k}\al_{l}},$}
\mbox{$c_{\bf 1}^{g} = \delta^{a_i a_j} \, \delta^{\al_k
  \al_l}, $}
\mbox{$c_{\bf 8S}^{g}=  T^b _{\alpha_l \alpha_k} d^{b a_i a_j} ,$}
\mbox{$c^{g}_{\bf 8A} = i T^b _{\alpha_l \alpha_k} f^{b a_i a_j} $}. 
In this basis the soft anomalous dimension matrix becomes diagonal in the absolute production threshold limit~\cite{KS}. However, for the invariant mass kinematics the soft anomalous dimension matrix with full kinematic dependence is required, which is not diagonal. The NLL resummed cross section in the
$N$-space has the form~\cite{Contopanagos:1996nh,Kidonakis:1998nf}
\begin{equation}
\label{eq:res:fact}
\frac{d\sigh^{{\rm (res)}}_{ij\tosv kl B,N}}{dQ^2} = \sum_{I,J} \,H_{ij\tosv kl B,IJ}(N)\,\tilde{S}_{ij\tosv kl B,JI}(N)\,\Delta^i_{N+1} \Delta^j_{N+1} 
\end{equation}
where we suppress explicit dependence on the scales. 
The indices  $I$ and $J$ in Eq.~(\ref{eq:res:fact}) indicate colour space matrix indices. The colour-channel-dependent hard contributions originating from the LO partonic cross sections in Mellin-moment space and at, higher orders, the contributions beyond LO and are denoted by $H_{ ij\tosv kl B,IJ}(N)$. The radiative factors
$\Delta^i_{N}$ describe the effect of the soft gluon radiation
collinear to the initial state partons and are universal, see e.g.~\cite{BCMNtop} . Large-angle
soft gluon emission is accounted for by the factors $S_{ ij\tosv kl B,IJ}(N)$ which are directly related to the soft gluon anomalous dimension calculated in~\cite{Kulesza:2015vda}. As indicated by the lower indices, the wide-angle soft emission depends on the partonic process under consideration and the colour configuration of the participating particles. In the limit of $p_B,m_B\to0$ the matrix $S_{ ij\tosv kl B,IJ}(N)$ coincides with the corresponding matrix for a $2 \to 2$ process $ij \to kl$. In addition the absolute threshold limit also reproduces the same matrix as for this $2 \to 2$ process~\cite{Kulesza:2015vda}. In our calculations we consider all perturbative functions governing the radiative factors up to the terms needed to obtain NLL accuracy in the resummed expressions. 

The function $S_{ ij\tosv kl B,IJ}(N)$ is given by~\cite{Kidonakis:1998nf}
\begin{eqnarray}
\tilde{S}_{ij\to kl}\left(N\right) & = & \bar{\mathrm{P}}\exp\left[\int_{\mu}^{Q/N}\frac{dq}{q}\Gamma_{ij\to kl}^{\dagger}\left(\alpha_{\mathrm{s}}\left(q^{2}\right)\right)\right]\tilde{S}_{ij\to kl}\nonumber \\
&  & \times\mathrm{P}\exp\left[\int_{\mu}^{Q/N}\frac{dq}{q}\Gamma_{ij\to kl}\left(\alpha_{\mathrm{s}}\left(q^{2}\right)\right)\right]\nonumber
\end{eqnarray}
where at the lowest order the matrix $\tilde{S}_{ij\to kl,IJ}=\mathrm{Tr}\left[c_I^\dagger c_J\right]$ and $\mathrm{P}$ and  $\bar{\mathrm{P}}$ denote the path- and reverse path-ordering in the variable $q$ respectively. If the soft anomalous dimension matrix is diagonal this expression simplifies, however this is not the case for the invariant mass threshold. Therefore we shall make use of the method of~\cite{Kidonakis:1998nf} in order to diagonalize the soft anomalous dimension matrix. Using this method we transform to a diagonal basis $R$
\begin{eqnarray}
\Gamma_{R} & = & R^{-1}\Gamma R\nonumber \\
H_{R} & = & R^{-1}H\left(R^{-1}\right)^{\dagger}\\
S_{R} & = & R^{\dagger}SR\nonumber 
\end{eqnarray}
In this basis we can write $S_{ ij\tosv kl B,R,IJ}(N)$ as
\begin{equation}
\tilde{S}_{ij\to kl,R,IJ}\left(N\right) = \tilde{S}_{ij\to kl,R,IJ}\exp\left[\int_{\mu}^{Q/N}\frac{dq}{q}\left\{ \lambda_{R,I}^{*}\left(\alpha_{\mathrm{s}}\left(q^{2}\right)\right)+\lambda_{R,J}\left(\alpha_{\mathrm{s}}\left(q^{2}\right)\right)\right\} \right]
\end{equation}
with $\lambda_{R,I}$ the eigenvalues of the matrix $\Gamma_{IJ}$. 

The matrix $H_{ ij\tosv kl B,IJ}(N)$ is described by the Born cross section projected onto the colour basis. However beyond NLL accuracy higher order terms in $H_{ ij\tosv kl B,IJ}(N)$ and $S_{ ij\tosv kl B,IJ}$ start contributing. These terms lead to the invariant mass resummation equivalent of the matching coefficient for absolute threshold resummation. In practice we split this term into the LO matrix in colour space and a single coefficient averaged over colour space for the higher order contributions
\begin{equation}
H_{ ij\tosv kl B,IJ}(N)=H^{(0)}_{ ij\tosv kl B,IJ}{C}_{ij\tosv klB}
\end{equation}
The coefficient ${C}_{ij\tosv klB}= 1 + \frac{\als}{\pi} {C}^{(1)}_{ij\tosv klB}+ \dots$
contains all non-logarithmic contributions to the NLO cross section taken in the invariant mass threshold limit. More specifically, these consist of the full virtual
corrections including the Coulomb corrections and $N$-independent non-logarithmic contributions from soft emissions. 
Although formally the coefficient $C_{ij\tosv kl B}$ begin to contribute at NNLL accuracy, in our numerical studies of the $pp \to \ttbh$ process we consider both the case of $C_{ij\tosv kl B}=1$, i.e. with the first-order corrections to the coefficients neglected, as well as the case with these corrections included. In the latter case we treat the Coulomb corrections and the hard contributions additively, i.e. 
${C}_{ij\tosv klB}^{(1)}={C}_{ij\tosv klB}^{(1, \rm hard)}+{C}_{ij\tosv klB}^{(1, \rm Coul)} .
$
For $k,l$ denoting massive quarks the Coulomb corrections are ${C}_{ij\tosv klB,{\bf 1}}^{(1, \rm Coul)} = \CF \pi^2 /(2 \beta_{kl})$ and  ${C}_{ij\tosv klB,{\bf 8}}^{(1, \rm Coul)} = (\CF -\CA/2) \pi^2 /(2 \beta_{kl})$ with $\beta_{kl}=\sqrt{1- 4m_t^2/\shat_{kl}}$ and $\shat_{kl}=(p_t+p_{\tb})^2$. As the $N$-independent non-logarithmic contributions from soft emission are accounted for using a modification of the techniques developed for the $2\to2$ case~\cite{Beenakker:2011sf,Beenakker:2013mva}, the problem of calculating the $C_{ij\tosv \ttbh}^{(1)}$ coefficient reduces to calculation of virtual corrections to the process. We extract them numerically using the publicly available POWHEG implementation of the $\ttbh$ process~\cite{Hartanto:2015uka}, based on the calculations developed in~\cite{Reina:2001sf}. The results are then cross-checked using the standalone MadLoop implementation in aMC@NLO~\cite{Hirschi:2011pa}. 

The resummation-improved NLO+NLL cross sections for the $pp \to \ttbh$ process are then
obtained through matching the NLL resummed expressions with 
the full NLO cross sections
\begin{eqnarray}
\label{hires}
&& \si^{\rm (NLO+NLL)}_{h_1 h_2 \tosv kl B}(\rho, \mu_F^2, \mu_R^2)\! =\! 
\si^{\rm (NLO)}_{h_1 h_2 \tosv kl B}(\rho,\mu_F^2, \mu_R^2) +   \si^{\rm
  (res-exp)}_
{h_1 h_2 \tosv kl B}(\rho, \mu_F^2, \mu_R^2) \nn \\
&&\!\!\!\!\!\!\!\!\!{\rm with} \nn \\
&& \si^{\rm
  (res-exp)}_{h_1 h_2 \tosv kl B}  \! =   \sum_{i,j}\,
\int_{\cal C}\,\frac{dN}{2\pi
  i} \; \rho^{-N} f^{(N+1)} _{i/h{_1}} (\mu_F^2) \, f^{(N+1)} _{j/h_{2}} (\mu_F^2) 
\nn \\ 
&& 
\! \times\! \left[ 
\sigh^{\rm (res)}_{ij\tosv kl B,N} (\mu_F^2, \mu_R^2)
-  \sigh^{\rm (res)}_{ij\tosv kl B,N} (\mu_F^2, \mu_R^2)
{ \left. \right|}_{\scriptscriptstyle({NLO})}\! \right], 
\end{eqnarray}
where $\sigh^{\rm (res)}_{ij\tosv
  kl B,N}$ is given in Eq.~(\ref{eq:res:fact}) and  $ \sigh^{\rm
  (res)}_{ij\tosv kl B,N} \left. \right|_{\scriptscriptstyle({NLO})}$ represents its perturbative expansion truncated at NLO.
The moments of the parton 
distribution functions (pdf) $f_{i/h}(x, \mu^2_F)$ are 
defined in the standard way 
$f^{(N)}_{i/h} (\mu^2_F) \equiv \int_0^1 dx \, x^{N-1} f_{i/h}(x, \mu^2_F)$. 
The inverse Mellin transform (\ref{hires}) is evaluated numerically using 
a contour ${\cal C}$ in the complex-$N$ space according to the ``Minimal Prescription'' 
method developed in Ref.~\cite{Catani:1996yz}.  

\section{Numerical predictions}
The numerical results presented in this section are obtained with $m_t=173$ GeV, $m_H=125$~GeV and MMHT14 pdf sets~\cite{Harland-Lang:2014zoa}. We choose the central renormalization and factorization scales as $\mu_{F, 0} =\mu_{R, 0}= m_t +m_H/2$, in accordance with~\cite{Dittmaier:2011ti}. The NLO cross section is calculated using the aMC@NLO code~\cite{Alwall:2014hca}.

\begin{figure}
\centering
\begin{tabular}{cc}
	\includegraphics[width=0.45\textwidth]{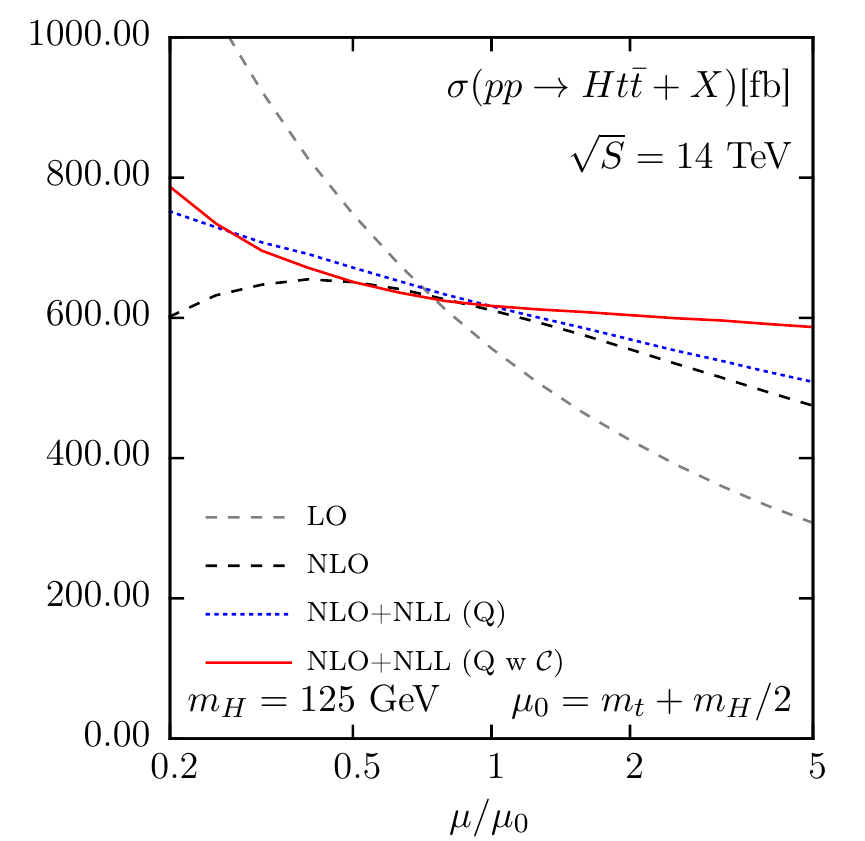} &
	\includegraphics[width=0.45\textwidth]{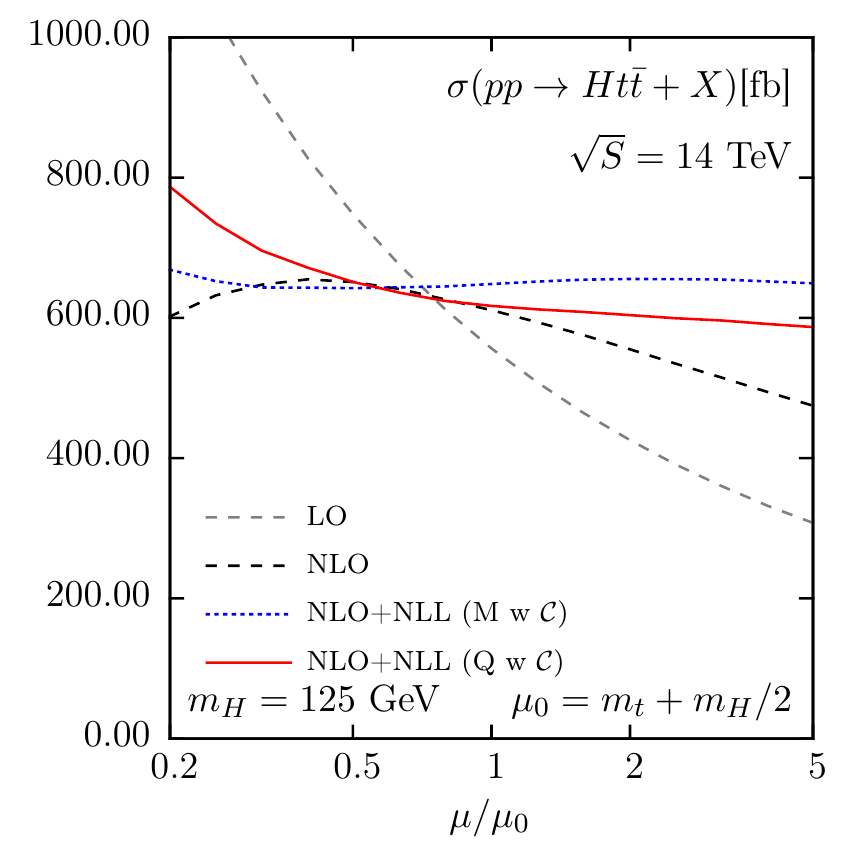} \tabularnewline
	\hspace{1.7em} (a) &\hspace{1.7em} (b) \tabularnewline
\end{tabular}
\caption{Scale dependence of the LO, NLO and NLO+NLL cross sections at $\sqrt S=14$ TeV LHC collision energy. The results are obtained while simultaneously varying $\mu_F$ and $\mu_R$, $\mu=\mu_F=\mu_R$.} 
\label{f:scaledependence:sim}
\end{figure}

In Figure~\ref{f:scaledependence:sim} we analyse the scale dependence of the resummed total cross section for $pp \to \ttbh$ at 14 TeV, varying simultaneously the factorization and renormalization scales, $\mu_F$ and $\mu_R$. In Figure~\ref{f:scaledependence:sim}~(a) a comparison between including and excluding the matching coefficient is shown for invariant mass resummation, where the inclusion of the matching coefficient is indicated by "w~$C$" and the use of invariant mass resummation is indicated by its scale $Q$. Whereas, Figure~\ref{f:scaledependence:sim}~(b) compares the previous result for absolute threshold resummation to the new result for invariant mass resummation, here absolute threshold resummation is indicated by its scale $M=2m_t+m_H$ and invariant mass resummation is again indicated by $Q$. 

Figure~\ref{f:scaledependence:sim}~(a) demonstrates that adding the soft gluon corrections and higher order hard contributions stabilizes the dependence on $\mu=\mu_F=\mu_R$ of the NLO+NLL predictions with respect to NLO.  The central values, calculated at $\mu=\mu_0= m_t +m_H/2$, and the scale uncertainty from simultaneous variation of the scales at $\sqrt S=14$ TeV changes from $613_{-9.4\%}^{+6.2\%}$ fb at NLO to  $619_{-2.4\%}^{+5.2\%}$ fb at NLO+NLL (with $C^{(1)}_{ij \tosv \ttbh}$ coefficients included). The increase in cross section for low scales can possibly be attributed to the fact that the $qg$ channel only begins to contribute at NLO and therefore does not undergo the resummation procedure and is not taken into account at higher orders. It is also clear from Figure~\ref{f:scaledependence:sim}~(a) that the coefficient $C_{ij\tosv \ttbh}^{(1)}$ strongly impact the predictions, especially at higher scales. In fact, their effect is more important than the effect of the logarithmic corrections alone for large scales. This observation also indicates the relevance of the contributions originating from the region away from the threshold which need to be known in order to further improve theoretical predictions.

In Figure~\ref{f:scaledependence:sim}~(b) it can be seen that there is a difference in the size of the correction for invariant mass threshold resummation and absolute threshold resummation. 

\begin{figure}
	\centering
	\includegraphics[width=0.45\textwidth]{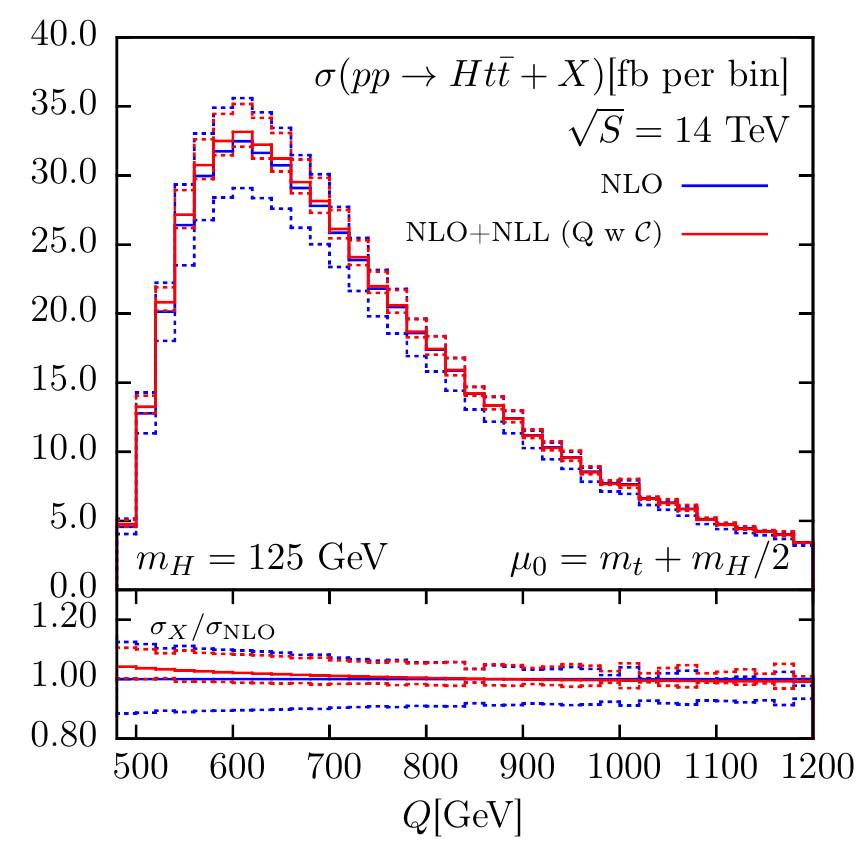}
	\caption{Invariant mass dependence of the NLO and NLO+NLL cross sections differential at $\sqrt S=14$ TeV LHC collision energy. The uncertainty bands are obtained by simultaneously varying $\mu_F$ and $\mu_R$ by a factor 2 around $\mu_0$. The lower half shows the ratio with respect to the central value of the NLO result.} 
	\label{f:Qdependence}
\end{figure}

Unlike absolute threshold resummation, invariant mass resummation allows differential distributions to be computed, specifically the invariant mass distribution. Figure~\ref{f:Qdependence} shows the  invariant mass distribution with the scale variation for simultaneous $\mu_R$ and $\mu_F$ variation. From this we can see that the invariant mass distribution is stable with respect to higher order soft gluon emission at the chosen central scale. At the hand of the increase in the size of the cross section for the lower uncertainty bound, which is taken at $2\;\mu_0$, it can be seen that for larger scale choices the corrections are significantly larger. An example of such a larger scale choice is the peak of the invariant mass distribution, $\mu\approx2.64\;\mu_0$  as is used in~\cite{Broggio:2015lya}.

 The effect of including NLL corrections is summarized in Table~\ref{t:results} for the LHC collision energies of 13 and 14 TeV.  Here we choose to estimate the theoretical uncertainty due to scale variation using the 7-point method, where the minimum and maximum values obtained with
 $(\mu_F/\mu_0, \mu_R/\mu_0) = (0.5,0.5), (0.5,1), (1,0.5), (1,1), (1,2), (2,1), (2,2)$ are considered. The invariant mass threshold NLO+NLL predictions  including the  $C^{(1)}_{ij \tosv t \tb H}$ show a significant reduction of the scale uncertainty, compared to NLO results. The reduction of the positive and negative scale errors amounts to around 20-30\% of the NLO error for $\sqrt S=13, 14$ TeV. If invariant mass threshold is compared to absolute threshold resummation the scale uncertainty is more greatly reduced and, as previously stated, the K-factor is smaller. The scale uncertainty of the predictions are still larger than the pdf uncertainty of the NLO predictions which is not expected to be significantly influenced by the soft gluon corrections.

\begin{table}
\begin{center}
\begin{tabular}{|c c c c c c|}
	\hline
	$\sqrt{S}$ {[}TeV{]} & NLO {[}fb{]} & \multicolumn{2}{c}{NLO+NLL M with $C$} & \multicolumn{2}{c|}{NLO+NLL Q with $C$} \tabularnewline
	& & Value {[}fb{]} & K-factor & Value {[}fb{]}  & K-factor \tabularnewline
	\hline 
	13 & $506_{-9.4\%}^{+5.9\%}$ & $537_{-5.5\%}^{+8.2\%}$ & 1.06 & $512_{-6.2\%}^{+5.1\%}$ & 1.01 \tabularnewline
	\hline 
	14 & $613_{-9.4\%}^{+6.2\%}$ & $650_{-5.7\%}^{+7.9\%}$ & 1.06 & $619_{-6.4\%}^{+5.2\%}$ & 1.01 \tabularnewline
	\hline 
\end{tabular}
\end{center}
\caption{NLO+NLL and NLO total cross sections for $pp \to \ttbh$ at $\sqrt S=13$ and 14 GeV. The NLO+NLL results are shown with $C$-coefficient and for both absolute threshold and invariant mass threshold. The error ranges given together with the NLO and NLO+NLL results indicate the scale uncertainty computed by use of the seven point method.}
\label{t:results}
\end{table}

\noindent
{\bf Acknowledgements}

This work has been supported in part by the German Research Foundation (DFG) grant KU 3103/1. Support of the Polish National Science Centre grants no.\ DEC-2014/13/B/ST2/02486 is gratefully acknowledged.
TS acknowledges support in the form of a scholarship of Marian Smoluchowski Research
Consortium Matter Energy Future from KNOW funding. This work is also partly supported by the U.S. National Science Foundation, under grant PHY--0969510, the LHC Theory Initiative.


\begin{thebibliography}{99}

\bibitem{Aad:2012tfa}
  G.~Aad {\it et al.}  [ATLAS Collaboration],
  Phys.\ Lett.\ B {\bf 716} (2012) 1
  [arXiv:1207.7214 [hep-ex]].

\bibitem{Chatrchyan:2012ufa}
  S.~Chatrchyan {\it et al.}  [CMS Collaboration],
  Phys.\ Lett.\ B {\bf 716} (2012) 30
  [arXiv:1207.7235 [hep-ex]].


\bibitem{Beenakker:2001rj}
  W.~Beenakker, S.~Dittmaier, M.~Kr\"amer, B.~Plumper, M.~Spira and P.~M.~Zerwas,
  Phys.\ Rev.\ Lett.\  {\bf 87} (2001) 201805
  [hep-ph/0107081];
  W.~Beenakker, S.~Dittmaier, M.~Kr\"amer, B.~Plumper, M.~Spira and P.~M.~Zerwas,
  Nucl.\ Phys.\ B {\bf 653} (2003) 151
  [hep-ph/0211352].


\bibitem{Reina:2001sf}
  L.~Reina and S.~Dawson,
  Phys.\ Rev.\ Lett.\  {\bf 87} (2001) 201804
  [hep-ph/0107101];
  L.~Reina, S.~Dawson and D.~Wackeroth,
  Phys.\ Rev.\ D {\bf 65} (2002) 053017
  [hep-ph/0109066];
  S.~Dawson, L.~H.~Orr, L.~Reina and D.~Wackeroth,
  Phys.\ Rev.\ D {\bf 67} (2003) 071503
  [hep-ph/0211438];
  S.~Dawson, C.~Jackson, L.~H.~Orr, L.~Reina and D.~Wackeroth,
  Phys.\ Rev.\ D {\bf 68} (2003) 034022
  [hep-ph/0305087].

\bibitem{Hirschi:2011pa}
  V.~Hirschi, R.~Frederix, S.~Frixione, M.~V.~Garzelli, F.~Maltoni and R.~Pittau,
  JHEP {\bf 1105} (2011) 044
  [arXiv:1103.0621 [hep-ph]].

\bibitem{Frederix:2011zi}
  R.~Frederix, S.~Frixione, V.~Hirschi, F.~Maltoni, R.~Pittau and P.~Torrielli,
  Phys.\ Lett.\ B {\bf 701} (2011) 427
  [arXiv:1104.5613 [hep-ph]].


\bibitem{Garzelli:2011vp}
  M.~V.~Garzelli, A.~Kardos, C.~G.~Papadopoulos and Z.~Trocsanyi,
  Europhys.\ Lett.\  {\bf 96} (2011) 11001
  [arXiv:1108.0387 [hep-ph]].

\bibitem{Hartanto:2015uka}
  H.~B.~Hartanto, B.~Jager, L.~Reina and D.~Wackeroth,
  Phys.\ Rev.\ D {\bf 91} (2015) 9,  094003
  [arXiv:1501.04498 [hep-ph]].

\bibitem{Frixione:2014qaa}
  S.~Frixione, V.~Hirschi, D.~Pagani, H.~S.~Shao and M.~Zaro,
  JHEP {\bf 1409} (2014) 065
  [arXiv:1407.0823 [hep-ph]].

\bibitem{Yu:2014cka}
  Y.~Zhang, W.~G.~Ma, R.~Y.~Zhang, C.~Chen and L.~Guo,
  Phys.\ Lett.\ B {\bf 738} (2014) 1
  [arXiv:1407.1110 [hep-ph]].

\bibitem{Frixione:2015zaa}
  S.~Frixione, V.~Hirschi, D.~Pagani, H.-S.~Shao and M.~Zaro,
  JHEP {\bf 1506} (2015) 184
  [arXiv:1504.03446 [hep-ph]].

\bibitem{Denner:2015yca}
  A.~Denner and R.~Feger,
JHEP {\bf 1511} (2015) 209
  [arXiv:1506.07448 [hep-ph]].
  
\bibitem{Kulesza:2015vda}
  A.~Kulesza, L.~Motyka, T.~Stebel and V.~Theeuwes,
  JHEP {\bf 1603} (2016) 065
  [arXiv:1509.02780 [hep-ph]].
  
\bibitem{Broggio:2015lya}
  A.~Broggio, A.~Ferroglia, B.~D.~Pecjak, A.~Signer and L.~L.~Yang,
  JHEP {\bf 1603} (2016) 124
  [arXiv:1510.01914 [hep-ph]].

\bibitem{Bonciani:2003nt}
  R.~Bonciani, S.~Catani, M.~L.~Mangano and P.~Nason,
  Phys.\ Lett.\ B {\bf 575} (2003) 268
  [hep-ph/0307035];

\bibitem{Aybat:2006wq}
  S.~M.~Aybat, L.~J.~Dixon and G.~F.~Sterman,
  Phys.\ Rev.\ Lett.\  {\bf 97} (2006) 072001
  [hep-ph/0606254];
  S.~M.~Aybat, L.~J.~Dixon and G.~F.~Sterman,
  Phys.\ Rev.\ D {\bf 74} (2006) 074004
  [hep-ph/0607309].

\bibitem{Contopanagos:1996nh}
  H.~Contopanagos, E.~Laenen and G.~F.~Sterman,
  Nucl.\ Phys.\ B {\bf 484} (1997) 303
  doi:10.1016/S0550-3213(96)00567-6
  [hep-ph/9604313].

\bibitem{KS}
  N.~Kidonakis and G.~Sterman,
  Phys.\ Lett.\  B {\bf 387} (1996) 867;
  Nucl.\ Phys.\  B {\bf 505}, 321 (1997).
  
\bibitem{Kidonakis:1998nf}
N.~Kidonakis, G.~Oderda and G.~F.~Sterman,
Nucl.\ Phys.\ B {\bf 531} (1998) 365
[hep-ph/9803241].

\bibitem{BCMNtop}
  R.~Bonciani, S.~Catani, M.~L.~Mangano and P.~Nason,
  Nucl.\ Phys.\  B {\bf 529}, 424 (1998).

\bibitem{Beenakker:2011sf}
  W.~Beenakker, S.~Brensing, M.~Kramer, A.~Kulesza, E.~Laenen and I.~Niessen,
  JHEP {\bf 1201} (2012) 076
  [arXiv:1110.2446 [hep-ph]].

\bibitem{Beenakker:2013mva}
  W.~Beenakker {\it et al.},
  JHEP {\bf 1310} (2013) 120
  [arXiv:1304.6354 [hep-ph]].


\bibitem{Catani:1996yz}
  S.~Catani, M.~L.~Mangano, P.~Nason and L.~Trentadue,
  Nucl.\ Phys.\  B {\bf 478}, 273 (1996).



\bibitem{Harland-Lang:2014zoa}
  L.~A.~Harland-Lang, A.~D.~Martin, P.~Motylinski and R.~S.~Thorne,
  Eur.\ Phys.\ J.\ C {\bf 75} (2015) 5,  204
  [arXiv:1412.3989 [hep-ph]].

\bibitem{Dittmaier:2011ti}
  S.~Dittmaier {\it et al.}  [LHC Higgs Cross Section Working Group Collaboration],
  arXiv:1101.0593 [hep-ph].


\bibitem{Alwall:2014hca}
  J.~Alwall {\it et al.},
  JHEP {\bf 1407} (2014) 079
  [arXiv:1405.0301 [hep-ph]].

\bibitem{Sterman:2013nya}
  G.~Sterman and M.~Zeng,
  JHEP {\bf 1405} (2014) 132
  doi:10.1007/JHEP05(2014)132
  [arXiv:1312.5397 [hep-ph]].

\end{thebibliography}
\end{document}